\documentclass[12pt]{iopart}
\usepackage{iopams}  
\usepackage{graphicx} 
\begin{document}

\title[Towards the metal-insulator transition in fluid alkali metals]
{Towards first-principles understanding of the metal-insulator transition 
in fluid alkali metals}

\author{H Maebashi and Y Takada}

\address{Institute for Solid State Physics, University of Tokyo, Kashiwa, 
Chiba 277-8581, Japan}
\ead{maebashi@issp.u-tokyo.ac.jp}
\begin{abstract}
By treating the electron-ion interaction as perturbation in the first-principles 
Hamiltonian, we have calculated the density response functions of a fluid alkali 
metal to find an interesting charge instability due to anomalous electronic 
density fluctuations occurring at some finite wave vector ${\bi Q}$ in a dilute 
fluid phase above the liquid-gas critical point. Since $|{\bi Q}|$ is smaller 
than the diameter of the Fermi surface, this instability necessarily impedes 
the electric conduction, implying its close relevance to the metal-insulator 
transition in fluid alkali metals.
\end{abstract}

\submitto{\JPCM}
\maketitle

\section{Introduction}
%%%%%%%%%%%%%% Paragraph 1-1: MIT in fluid alkali metals           %%%%%%%%%%%%%
The metal-insulator transition in fluid alkali metals such as Rb and Cs has long 
been attracting attention \cite{Mott90,Hensel98}, partly because this might be 
a faithful manifestation of the ``Mott transition" in its original sense 
\cite{Mott49} or the one driven by an insufficiently-screened Coulomb potential 
and partly because this accompanies the liquid-gas phase transition, the same 
situation as first considered in mercury by Landau and Zeldovitch \cite{LZ43}. 

%%%%%%%%%%%%%% Paragraph 1-2: definition of a fluid phase          %%%%%%%%%%%%%
In the density-temperature phase diagram of the alkali metals, the liquid-phase 
region is bounded above by the critical point and below by the triple point, 
as in the systems of rare-gas atoms \cite{HM05}. Above the critical point, 
there exists the fluid-phase region, an interesting phase in which the 
metal-insulator transition occurs. In many respects, the properties of the 
dense supercritical fluid are not very different from those of the liquid 
and a well-known charge instability develops near the transition from liquid to 
solid, as the electronic (ionic) density, $n$, increases or as the Wigner-Seitz 
radius of electrons (ions) in atomic units, $r_s$, decreases. We can expect, 
however, to observe a different type of charge instability closely related 
to the metal-insulator transition as $r_s$ increases in the fluid phase. 

%%%%%%%%%%%%%%%% Paragraph 1-3: Method in this paper    %%%%%%%%%%%%%%%%%%%%%%%%
Conventionally, liquid alkali metals are studied by treating the bare 
electron-ion (pseudo-)potential as perturbation to provide an effective 
ion-ion interaction via the sea of valence electrons described by the 
three-dimensional electron gas (3DEG) \cite{AS78}. The exchange-correlation 
effects of the valence electrons play an important role in the dilute fluid 
with large $r_s$. It is fortunate that in developing the first-principles 
calculation based on the density functional theory, useful information on 
3DEG is now available in a wide range of $r_s$. In this paper, 
we investigate the density response of the fluid alkali metal to find the 
charge instability for low densities by use of the standard perturbation 
approach, together with this information on 3DEG. Throughout this paper, 
we employ atomic units.  
   
\section{Response Functions in the First-Principles Hamiltonian \label{sec.2}}
%%%%%%%%%%%%%%%% Paragraph 2-1: Hamiltonian            %%%%%%%%%%%%%%%%%%%%%%%%%
The alkali metals composed of $N$ electrons and $N$ ions are well described 
by the following Hamiltonian $H$ as
%\numparts
\begin{equation}
H = T_{\rm e} + T_{\rm i} + U_{\rm ee} + U_{\rm ii} + U_{\rm ei}\,,
\label{sec.2: Hamiltonian}
\end{equation}
with
\begin{eqnarray}
T_{\rm e} = \sum_{j = 1}^{N} \frac{{\bi p}_j^2}{2m}, \qquad
T_{\rm i} = \sum_{j = 1}^{N} \frac{{\bi P}_j^2}{2M}, \qquad
 U_{\rm ee} = \frac{1}{2}\sum_{j \neq j'} V_{\rm ee}({\bi r}_j - {\bi r}_{j'}),
\cr
 U_{\rm ii} = \frac{1}{2}\sum_{j \neq j'} V_{\rm ii} ({\bi R}_j - {\bi R}_{j'}),
 \qquad
 U_{\rm ei} = \sum_{j,j'} V_{\rm ei}({\bi r}_j - {\bi R}_{j'}) \,,
\end{eqnarray}
%\endnumparts
where $\{ {\bi r}_j \}$ and $\{ {\bi R}_j \}$ denote, respectiviely, electronic 
and ionic coordinates, $\{ {\bi p}_j \}$ and $\{ {\bi P}_j \}$ corresponding 
momenta, $m$ and $M$ the masses, $V_{\rm ee} ({\bi r})$, $V_{\rm ii} ({\bi R})$ 
and  $V_{\rm ei}({\bi r})$ are the bare electron-electron, ion-ion and 
electron-ion interactions. At long distances $V_{\rm ii}({\bi R})$ as well as 
$V_{\rm ei}({\bi r})$ is represented by a purely Coulombic form, but it 
deviates from it at short-range distances due to the van der Waals attraction 
and the Born-Mayer repulsion between ionic cores for $V_{\rm ii}({\bi R})$ 
or due to orthogonality between the valence- and core-electron wave functions 
for $V_{\rm ei}({\bi r})$; the former contribution is weak for alkali metals 
and the latter can be well captured by adopting a suitable (local) 
pseudopotential $V_{\rm ei}({\bi r})$ $=$ $V_{\rm ps}({\bi r})$. In general, 
the Fourier transforms of these interactions $V_{\alpha \beta}({\bi q})$ can 
be written in terms of an effective valence, $Z_{\alpha}({\bi q})$, 
characterizing the long-range Coulomb interaction and a short-range part, 
$U_{\alpha \beta}({\bi q})$, as
\begin{equation}
V_{\alpha \beta}({\bi q}) = Z_{\alpha}({\bi q})Z_{\beta}({\bi q})v({\bi q}) 
+ U_{\alpha \beta}({\bi q})\,,
\label{sec.2: bare interaction}
\end{equation}
where $\alpha$ and $\beta$ denote species of the particles, e (electron) or 
i (ion). Here $Z_{\rm e}({\bi q})$ $=$ $-1$, $Z_{\rm i}({\bi q})$ $=$ 
$|V_{\rm ps}({\bi q})|/v({\bi q})$, $U_{\rm \alpha\beta}({\bi q})$ $=$ $0$ 
unless $\alpha$ $=$ $\beta$ $=$ i, and $U_{\rm ii}({\bi q})$ represents the 
short-range interaction between ionic cores. Reflecting the purely Coulombic 
nature of both $V_{\rm ii}({\bi R})$ and $V_{\rm ei}({\bi r})$ at long 
distances, $Z_{\rm i}({\bi q}) \to 1$ and $U_{\rm ii}({\bi q})$ converges to 
a finite value in the limit of $q \equiv |{\bi q}| \to 0$. 

%%%%%%%%% Paragraph 2-2: charge- and number density response function   %%%%%%%%
Since the first term in the right-hand side of \eref{sec.2: bare interaction} 
is proportional to $v({\bi q}) = 4 \pi/q^2$ and gets singular at $q \to 0$, 
it is useful to write the partial number-density response function 
$\chi_{\alpha\beta}({\bi q},\omega)$ in terms of its proper part 
$\Pi_{\alpha \beta}({\bi q}, \omega)$, which is irreducible with respect 
to $v({\bi q})$ but not to $U_{\alpha \beta}({\bi q})$, as
\begin{equation}
\fl \chi_{\alpha\beta}({\bi q},\omega) = - \Pi_{\alpha \beta}({\bi q}, \omega) 
- v({\bi q}) \sum_{\gamma,\delta = {\rm e}, {\rm i}}
 \Pi_{\alpha \gamma}({\bi q}, \omega) Z_{\gamma}({\bi q}) Z_{\delta}({\bi q}) 
\chi_{\delta \beta}({\bi q},\omega)\,,
 \label{sec.2: irreducible part}
\end{equation}
which is rewritten as
\begin{equation}
\fl
\chi_{\alpha\beta}({\bi q},\omega) = - \Pi_{\alpha \beta}({\bi q}, \omega) 
+ v({\bi q}) \sum_{\gamma,\delta = {\rm e}, {\rm i}} 
\frac{\Pi_{\alpha \gamma}({\bi q}, \omega)Z_{\gamma}({\bi q}) 
Z_{\delta}({\bi q}) \Pi_{\delta \beta}({\bi q},\omega) }{
 1 + v({\bi q}) \Pi_{ZZ}({\bi q}, \omega) }\,, 
 \label{sec.2: partial response function}
\end{equation}
where the polarization function is defined as $\Pi _{ZZ}({\bi q}, \omega)$ 
$\equiv$ $\sum_{\alpha \beta}$ $Z_{\alpha}({\bi q}) Z_{\beta}({\bi q}) 
\Pi_{\alpha \beta}({\bi q},\omega)$. In the two-component Coulomb system with 
the effective valences $Z_{\alpha}({\bi q})$, the charge density is given by 
the sum of ionic and electronic valence densities, while the (total) 
number density by the sum of ionic and electronic number densities. 
Then, by use of \eref{sec.2: partial response function}, the charge-density 
response function $\chi_{ZZ}({\bi q},\omega)$ $\equiv$ $\sum_{\alpha \beta}
Z_{\alpha}({\bi q}) Z_{\beta}({\bi q}) \chi_{\alpha \beta}({\bi q},\omega)$, 
the number-density response function $\chi_{NN}({\bi q},\omega)$ $\equiv$ 
$\sum_{\alpha \beta} \chi_{\alpha \beta}({\bi q},\omega)$, and the 
cross response function $\chi_{NZ}({\bi q},\omega)$ $\equiv$ 
$\sum_{\alpha \beta}Z_{\beta}({\bi q}) \chi_{\alpha \beta}({\bi q},\omega)$ 
are, respectively, obtained as 
\begin{eqnarray}
\chi_{ZZ}({\bi q},\omega)  
& =  - \frac{\Pi_{ZZ} ({\bi q}, \omega) }{1 + v({\bi q}) 
\Pi_{ZZ} ({\bi q}, \omega) }\,,
\label{sec.2: charge-density response function}
\\
\chi_{NN}({\bi q},\omega) 
&=  - \Pi_{NN} ({\bi q}, \omega)  +  \frac{v({\bi q}) 
\Pi_{NZ} ({\bi q}, \omega) \Pi_{ZN} ({\bi q}, \omega) 
}{1 + v({\bi q}) \Pi_{ZZ} ({\bi q}, \omega) }\,,
\label{sec.2: number-density response function}
\\
\chi_{NZ}({\bi q},\omega) 
& =  - \frac{\Pi_{NZ} ({\bi q}, \omega) }{1 + v({\bi q}) 
\Pi_{ZZ} ({\bi q}, \omega) }\,,
\label{sec.2: cross response function}
\end{eqnarray} 
where $\Pi _{NN}({\bi q}, \omega)$ $=$ $\sum_{\alpha \beta}$ 
$\Pi_{\alpha \beta}({\bi q},\omega)$, $\Pi _{NZ}({\bi q}, \omega)$ $=$ 
$\sum_{\alpha \beta}$ $Z_{\beta}({\bi q}) \Pi_{\alpha \beta}({\bi q},\omega)$, 
and $\Pi _{ZN}({\bi q}, \omega)$ $=$ $\sum_{\alpha \beta}$ $Z_{\alpha}({\bi q}) 
\Pi_{\alpha \beta}({\bi q},\omega)$.

%%%%%%% Paragraph 2-3: charge and compressible instabilities         %%%%%%%%%%%
Some comments are in order for the so-called $q$-limits of these response 
functions. In 3DEG or an unperturbed electron system for 
\eref{sec.2: Hamiltonian}, the charge neutrality condition and the 
compressibility sum rule lead, respectively, to the relations $\lim_{q \to 0}
\chi_{\rm ee}^{(0)}({\bi q},0) = 0$ and $\lim_{q \to 0}
\Pi_{\rm ee}^{(0)}({\bi q},0) = n^2 \kappa$ with the superscript 
$(0)$ implying ``unperturbed'' with respect to $U_{\rm ei}$, where $\kappa$ is 
the compressibility of 3DEG. These relations reflect a special feature of a 
single-component Coulomb system with a rigid compensated-charge background 
such as 3DEG in which we cannot distinguish the charge-density response 
function from the number-density one. In the two-component Coulomb system, 
on the other hand, there is a difference between them and the difference 
leads to important modifications on these relations. More specifically, 
the charge neutrality condition leads to 
\begin{equation}
\lim_{q \to 0}\chi_{ZZ}({\bi q},0) = \lim_{q \to 0} \chi_{NZ}({\bi q},0) = 0\,,
\label{sec3: charge neutrality condition}
\end{equation}
while the compressibility sum rule relates the $q$-limit of 
$\chi_{NN}({\bi q},\omega)$ to the isothermal compressibility 
$K_T $ of the total electron-ion system. Then, from \eref{sec.2: partial 
response function} and \eref{sec.2: number-density response function}, 
it is not hard to see that $\chi_{\alpha \beta}({\bi q},\omega)$ satisfies 
the following relations for arbitrary $\alpha$ and $\beta$ \cite{WH73}:
\begin{equation}
\lim_{q \to 0} \chi_{\alpha \beta}({\bi q},0) = \frac{1}{4} \lim_{q \to 0} 
\chi_{NN}({\bi q},0) = - n^2 K_T\,. 
\label{sec3: compressibility sum rule}
\end{equation}
Note that the $q$-limit of $\chi_{\rm ee}({\bi q},\omega)$ does not vanish 
but takes a finite value $- n^2 K_T$, although $\lim_{q \to 0}
\chi_{\rm ee}^{(0)}({\bi q},0) = 0$. 
By \eref{sec3: compressibility sum rule}, we can identify the liquid-gas 
phase transition from a singularity in 
$\chi_{\alpha \beta}({\bi q},0)$, particularly in the ionic number-density 
response function $\chi_{\rm ii}({\bi q},0)$ at $q = 0$, because $K_T$ 
diverges at the critical point. Note also that 
by \eref{sec3: charge neutrality condition}, the singularity implying a 
charge instability can occur only at a {\it finite} $q$.  

\section{Approximate Response Functions Based on Effective Ion-Ion Interaction}
%%%%%%%%%%%%%%% Paragraph 3-1: perturbation theory         %%%%%%%%%%%%%%%%%%%%%
Following Ashcroft and Stroud~\cite{AS78}, we can derive an effective 
ionic Hamiltonian $\tilde{H}_{\rm i} = T_{\rm i} + \tilde{U}_{\rm ii} + 
N \tilde{u}_0$ by applying a perturbative method with respect to $U_{\rm ei}$ 
in \eref{sec.2: Hamiltonian}, where $\tilde{u}_0$ is an energy shift 
independent of $\{ {\bi P}_j \}$ and $\{ {\bi R}_j \}$ (but dependent on $n$).  
The interaction term $\tilde{U}_{\rm ii}$ can be described by a pairwise 
sum of effective ion-ion interactions as
\begin{equation}
\tilde{U}_{\rm ii} = 
\frac{1}{2}\sum_{j \neq j'} 
\tilde{V}_{\rm ii}({\bf R}_j - {\bf R}_{j'})\,.
\end{equation}
Here the Fourier transform of the effective ion-ion interaction 
$\tilde{V}_{\rm ii}({\bf R})$ is given by
\begin{equation}
\tilde{V}_{\rm ii}({\bi q}) = Z_{\rm i}({\bi q})^2v({\bi q})/
\varepsilon ({\bi q}) + U_{\rm ii}({\bi q})\,,
\label{sec.3: effective interaction}
\end{equation}
with $\varepsilon({\bi q})$ the static dielectric function of 3DEG defined as 
\begin{equation}
\varepsilon ({\bi q}) = 1 + v({\bi q}) \Pi_{\rm ee}^{(0)} ({\bi q},0)\,.
\end{equation}
\Eref {sec.3: effective interaction} indicates that the long-range part of 
the bare ion-ion interaction $V_{\rm ii}({\bi q})$, the first term in the 
right-hand side of \eref{sec.2: bare interaction}, has been screened by the 
surrounding valence electrons, while the short-range part $U_{\rm ii}({\bi q})$ 
remains intact.

%%%%%%%%%%%%%%%%% Paragraph 3-2: response functions         %%%%%%%%%%%%%%%%%%%%
We shall calculate the ionic structure factor $S_{\rm ii}({\bi q})$ in this 
electron-ion system by using the effective ionic Hamiltonian. Since ions 
can be considered as classical particles, the classical version of the 
fluctuation-dissipation theorem relates $S_{\rm ii}({\bi q})$ to the static 
response function $\chi_{\rm ii}({\bi q},0)$ as
\begin{equation}
\chi_{\rm ii}({\bi q},0) =  - n S_{\rm ii}({\bi q})/T \,.
\label{sec.3: fluctuation-dissipation theorem}
\end{equation}
In the same spirit as in \eref{sec.2: irreducible part}, we shall introduce 
$\Pi_{\rm ii}({\bi q},0)$, the proper part irreducible with respect to 
$v({\bi q})$, to write $\chi_{\rm ii}({\bi q},0)$ as 
\begin{eqnarray}
\chi_{\rm ii}({\bi q},0) &= 
- [\, \Pi_{\rm ii}({\bi q},0)^{-1} + Z_{\rm i}({\bi q})^2 v({\bi q}) 
/\varepsilon ({\bi q}) \,]^{-1}
\cr
&= - \Pi_{\rm ii}({\bi q},0)  + \frac{ v({\bi q}) Z_{\rm i}({\bi q})^2 
\Pi_{\rm ii}({\bi q},0)^2}{
1 + v({\bi q}) [\,\Pi_{\rm ee}^{(0)}({\bi q},0) + Z_{\rm i}({\bi q})^2 
\Pi_{\rm ii}({\bi q},0)\,]}\,.
\label{sec.3: static response function}
\end{eqnarray}
By comparing \eref{sec.3: static response function} with 
\eref{sec.2: partial response function}, we find that 
$\Pi_{\rm ee}({\bi q},0)$ $=$ $\Pi_{\rm ee}^{(0)}({\bi q},0)$ and 
$\Pi_{\rm ei}({\bi q},0)$ $=$ $\Pi_{\rm ie}({\bi q},0) = 0$ at least 
in the present approximation. Substituting these equations and 
\eref{sec.3: fluctuation-dissipation theorem} into 
\eref{sec.2: partial response function}, we obtain 
\begin{equation}
\chi_{\rm ee}({\bi q},0) = 
- \frac{1}{v({\bi q})} \left( 1 - \frac{1}{\varepsilon ({\bi q})} \right)
- Z_{\rm i}({\bi q})^2 \left( 1 - \frac{1}{\varepsilon ({\bi q})} \right)^2
n S_{\rm ii}({\bi q})/T\,,
\label{sec.3: electron-electron response function}
\end{equation}
\begin{equation}
\chi_{\rm ei}({\bi q},0) = \chi_{\rm ie}({\bi q},0) =
- Z_{\rm i}({\bi q}) \left( 1 - \frac{1}{\varepsilon ({\bi q})} \right)
n S_{\rm ii}({\bi q})/T\,.
\label{sec.3: electron-ion response function}
\end{equation}
From \eref{sec.3: fluctuation-dissipation theorem}, 
\eref{sec.3: electron-electron response function} and 
\eref{sec.3: electron-ion response function}, the static 
charge-density, number-density and cross response functions 
are, respectively, given by 
\begin{equation}
\fl
\chi_{ZZ}({\bi q},0) = 
- \frac{1}{v({\bi q})} \left( 1 - \frac{1}{\varepsilon ({\bi q})} \right)
- \frac{Z_{\rm i}({\bi q})^2}{\varepsilon ({\bi q})^2} 
n S_{\rm ii}({\bi q})/T\,,
\end{equation}
\begin{equation}
\fl
\chi_{NN}({\bi q},0) = 
- \frac{1}{v({\bi q})} \left( 1 - \frac{1}{\varepsilon ({\bi q})} \right)
- \left( 1 + Z_{\rm i}({\bi q}) - \frac{Z_{\rm i}({\bi q})}
{\varepsilon ({\bi q})} \right)^2
n S_{\rm ii}({\bi q})/T\,,
\end{equation}
\begin{equation}
\fl
\chi_{NZ}({\bi q},0) = 
\frac{1}{v({\bi q})} \left( 1 - \frac{1}{\varepsilon({\bi q})} \right)
- \frac{Z_{\rm i}({\bi q})}{\varepsilon ({\bi q})}
\left( 1 + Z_{\rm i}({\bi q}) - \frac{Z_{\rm i}({\bi q})}
{\varepsilon ({\bi q})} \right)
n S_{\rm ii}({\bi q})/T\,.
\end{equation}
We can easily check that these response functions satisfy the required 
conditions of \eref{sec3: charge neutrality condition} and 
\eref{sec3: compressibility sum rule} with $\lim_{q \to 0} 
\chi_{\rm ii}({\bi q},0)$ $=$ $- n S_{\rm ii}({\bf 0})/T$ $=$ $-n^2 K_T$.

%%%%%%%%%%%%%%%%%%% Paragraph 3-3: Further simplification          %%%%%%%%%%%%%
In what follows, for simplicity, we take $V_{\rm ii} ({\bi R})$ $=$ 
$|{\bi R}|^{-1}$ and $V_{\rm ei} ({\bi r})$ $=$ 
$V_{\rm ps}^{({\rm A})}({\bi r})$, where $V_{\rm ps}^{({\rm A})}({\bi r})$ 
is the Ashcroft's empty-core pseudopotential, given by
\begin{equation}
V_{\rm ps}^{({\rm A})}({\bi r}) = 
\left\{
\begin{array}{ll}
0 & \quad |{\bi r}| < r_{\rm c} \\
-|{\bi r}|^{-1} & \quad |{\bi r}| > r_{\rm c}
\end{array}
\right.
\end{equation}
with $r_{\rm c}$ being the radius of the ionic core; this simplification 
leads to $Z_{\rm i}({\bi q})$ $=$ $\cos (q r_{\rm c})$ and 
$U_{\rm ii}({\bi q})$ $=$ $4 \pi [1 - \cos^2(q r_{\rm c})] /q^2$. 
For $\varepsilon({\bi q})$, we use an accurate parameterization of the 
diffusion Monte Carlo data of 3DEG at $T =0$ by Moroni {\it et al}. \cite{MCS95}. 

\section{Comparison of Mean-Field Approximation with Monte Carlo Simulation}
%%%%%%%%% Paragraph 4-1:  Methods in theory of simple liquids        %%%%%%%%%%%
We shall now solve the effective ionic Hamiltonian to obtain 
$S_{\rm ii}({\bi q})$ in the mean-field approximation, one of the standard 
methods in the theory of simple liquids, together with Monte Carlo simulations. 

%%%%%%%%%%%%%% Paragraph 4-2:  Mean-field approximation        %%%%%%%%%%%%%%%%%
Suppose that the effective ion-ion interaction, $\tilde{V}_{\rm ii}({\bi R})$ 
$=$ $(2 \pi^2 R)^{-1}\int_0^{\infty} {\rm d}q \tilde{V}_{\rm ii}(q) q \sin(q R)$ 
with $R \equiv |{\bf R}|$,  can be expressed as the sum of a ``reference" part, 
$v_0 ({\bi R})$, and a ``perturbation", $w ({\bi R})$, by
\begin{equation}
{\tilde V}_{\rm ii}({\bi R}) 
= v_0({\bi R}) + w ({\bi R})\,.
\end{equation}
In the mean-field approximation, the free-energy functional $F[n_{\rm i}]$ of 
the ionic number density $n_{\rm i}({\bi R})$ for the system of interest, 
characterized by the full potential ${\tilde V}_{\rm ii}({\bi R})$, 
is simply related to that of the reference system $F_0[n_{\rm i}]$ by
\begin{equation}
F[n_{\rm i}] =  
F_0[n_{\rm i}] + 
\frac{1}{2}\int\int n_{\rm i}({\bi R})
w ({\bi R}-{\bi R}') n_{\rm i}({\bi R}') 
{\rm d} {\bi R} {\rm d} {\bi R}'\,.
\end{equation}
Then, we can expand the thermodynamic potential $\Omega [n_{\rm i}]$ $=$ 
$F[n_{\rm i}]$ $-$ $\mu_{\rm i} \int n_{\rm i} ({\bi R}) {\rm d} {\bi R}$ 
(with $\mu_{\rm i}$ being the chemical potential of ions) up to second order 
with respect to the ionic number-density fluctuations $n_{\rm i}({\bi q})$ as
\begin{equation}
\Omega [n_{\rm i}] = \Omega(n) + 
\frac{T}{2 n}\sum_{{\bi q} \neq {\bf 0}} 
S_{\rm ii}({\bi q})^{-1} n_{\rm i}({\bi q)} n_{\rm i}(-{\bi q})
+ \cdots \,.
\end{equation}
Here $S_{\rm ii}({\bi q})$ is the ionic structure factor in the mean-field 
approximation, given by 
\begin{equation}
S_{\rm ii}({\bi q}) = [S_0({\bi q})^{-1} + n w({\bi q})/T ]^{-1}\,,
\label{sec.4: ionic structure factor}
\end{equation}
where $S_0({\bi q})$ is the structure factor of the reference system and 
$w({\bi q})$ is the Fourier transform of $w({\bi R})$.

\begin{figure}[b]
\begin{center}
\includegraphics[width=7.8cm, keepaspectratio]{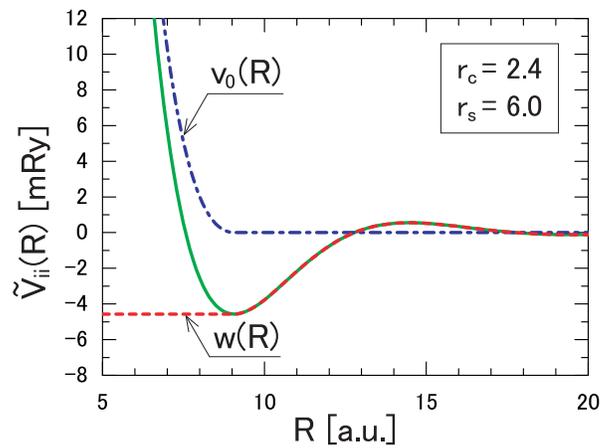}
\end{center}
\caption{Effective ion-ion interaction ${\tilde V}_{\rm ii}({\bi R})$ for 
$r_s = 8.0$ and $r_{\rm c} = 2.4$ (the solid curve), which is divided into 
the reference part, $v_0({\bi q})$, (the chain curve) and the 
perturbation, $w({\bi q})$, (the broken curve) in the WCA separation.}
\label{fig: WCA separation}
\end{figure}

%%%%%%%%%%%%%%%%%%%% Paragraph 4-3:  WCA separation        %%%%%%%%%%%%%%%%%%%%%
The result of the present mean-field approximation depends on how to separate 
the interaction into a reference part $v_0({\bi R})$ and a perturbation 
$w({\bi R})$. A number of separations have been proposed so far for the 
Lennard-Jones potential; among them, here we adopt the manner of Weeks, Chandler 
and Andersen, usually called the WCA separation \cite{WCA71}. In this separation, 
as shown in \fref{fig: WCA separation}, the interaction is split at $R = R_0$, 
the position of the minimum of ${\tilde V}_{\rm ii}({\bi R})$, into its purely 
repulsive and almost attractive parts; $v_{0}({\bi R}) = 
{\tilde V}_{\rm ii}({\bi R}) - E_0$ and $w({\bi R}) = E_0$ for $R < R_0$, 
while $v_{0}({\bi R}) = 0$ and $w({\bi R}) = {\tilde V}_{\rm ii}({\bi R})$ 
for $R > R_0$, where $E_0$ is the minimum value of ${\tilde V}_{\rm ii}({\bi R})$ 
at $R = R_0$. The reference system with the purely repulsive interaction 
$v_0({\bi R})$ can be effectively treated as a hard-sphere fluid where the 
diameter of the hard sphere is defined as 
\begin{equation}
d = \int_0^\infty ( 1 - {\rm e}^{-v_0({\bi R})/T } ){\rm d} R\,.
\end{equation}
Using the analytic solution of the Percus-Yevick equation for this hard-sphere 
fluid \cite{PY58,Thiele63,Wertheim63}, we can obtain $S_0({\bi q})$ and 
therefore $S_{\rm ii}({\bi q})$ by \eref{sec.4: ionic structure factor}. 

\begin{figure}[b]
\begin{center}
\includegraphics[width=7.8cm, keepaspectratio]{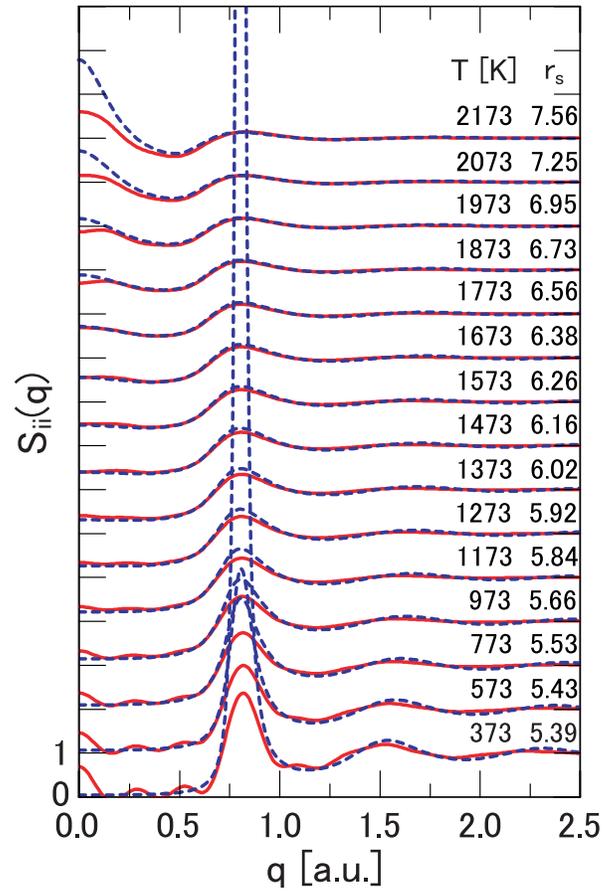}
\end{center}
\caption{Ionic structure factor $S_{\rm ii}({\bi q})$ in a liquid phase 
of Rb ($r_{\rm c} = 2.4$). The solid and broken lines represent Monte-Carlo 
and mean-field results, respectively.}
\label{fig: MFA vs MCS}
\end{figure}

%%%%%%%%%%%%%%%%%%%% Paragraph 4-4:  MFA vs MCS         %%%%%%%%%%%%%%%%%%%%%%%%
\Fref{fig: MFA vs MCS} shows the ionic structure factor $S_{\rm ii}({\bi q})$ 
calculated in this mean-field approximation based on the WCA separation (the 
broken curves) and by Monte Carlo simulations (the solid curves) 
for temperatures and densities corresponding to the conditions of the recent 
experiments for a liquid Rb \cite{MTI07}. Note that for all the values of $T$ 
and $r_s$ in \fref{fig: MFA vs MCS} the Monte Carlo results are in good accord 
with the experimental data, which has not been shown here. Although the 
quantitative disagreement between mean-field and Monte-Carlo results can be 
observed in the vicinity of the triple point (e.g. $T$ $=$ 373K, $r_s = 5.39$) 
and in the vicinity of the critical point (e.g. $T$ $=$ 2173K, $r_s = 7.56$), 
we can find that the mean-field approximation reproduces at least qualitative 
features of the Monte-Calro results for $S_{\rm ii}({\bi q})$ such as the peak 
position. In the next section, with this knowledge of the mean-field 
approximation, we will investigate the charge-density response of the system 
in a fluid phase above the critical point. 

\section{Charge Instability in a Fluid Phase above the Critical Point}
\begin{figure}[b]
\begin{center}
\includegraphics[width=7.8cm, keepaspectratio]{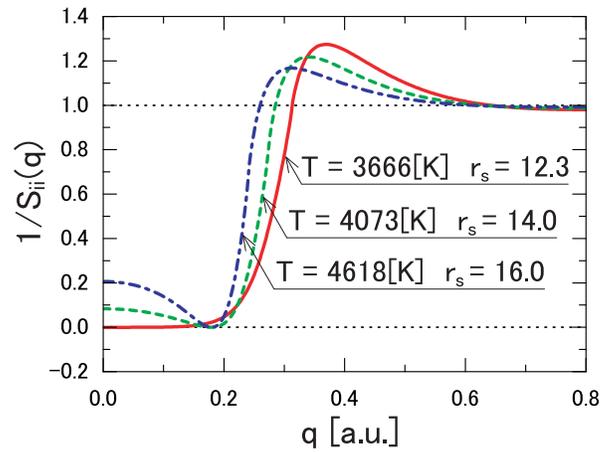}
\end{center}
\caption{Inverse of the ionic structure factor $S_{\rm ii}({\bi q})$ in the 
mean-field approximation based on the WCA separation for low densities of 
$r_s = 12.3$, $14.0$ and $16.0$ with the core radius of $r_{\rm c} = 2.4$.}
\label{fig: structure factor}
\end{figure}

\begin{figure}[b]
\begin{center}
\includegraphics[width=7.8cm, keepaspectratio]{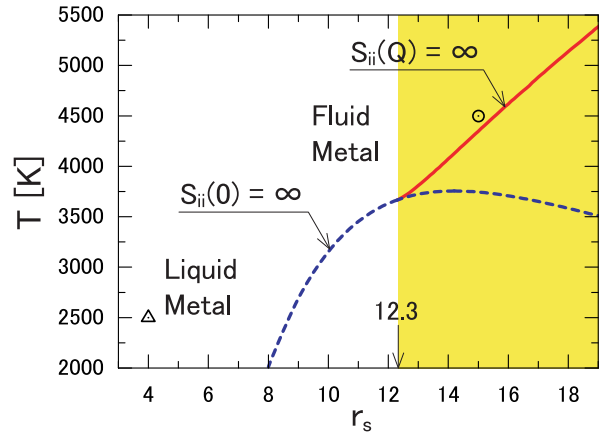}
\end{center}
\caption{Charge and compressible instabilities in the mean-field approximation 
based on the WCA separation for $r_{\rm c} = 2.4$. The structure factor 
$S_{\rm ii}({\bi q})$ diverges at ${\bi q} = {\bi Q}$ (with ${\bi Q}$ beeing 
a certain finite momentum) along the solid curve, while it does at ${\bi q} = 
{\bf 0}$ along the broken curve in the $r_s$-$T$ plane.}
\label{fig: phase diagram}
\end{figure}

%%%%%%%%% Paragraph 5-1: Divergence in the structure factor    %%%%%%%%%%%%%%%%%
As discussed in \sref{sec.2}, a charge instability is signaled by a divergence 
in $\chi_{\rm ii}({\bi q},0)$ $=$ $- n S_{\rm ii}({\bi q})/T$ at $q = Q 
\equiv |{\bi Q}|$ with ${\bi Q}$ being a certain {\it finite} wave vector, 
while the compressible instability by that at $q = 0$. In fact the mean-field 
approximation based on the WCA separation leads us to the result that 
$S_{\rm ii}({\bi q})$ diverges for low densities as shown in 
\fref{fig: structure factor}, from which the divergence is seen to occur 
at $q = Q \sim 0.2$ for $r_s = 14$ (the broken curve) and $r_s = 16$ (the chain 
curve), but at $q =0$ for $r_s = 12.3$ (the solid curve). 
\Fref{fig: phase diagram} presents the ``phase diagram'' in the $r_s$-$T$ 
plane in which the charge instability occurs along the solid curve, while the 
compressible instability does along the broken curve, the former preceding 
the latter with decreasing temperature in the shaded region of $r_s > 12.3$. 

\begin{figure}[b]
\begin{center}
\includegraphics[width=7.8cm, keepaspectratio]{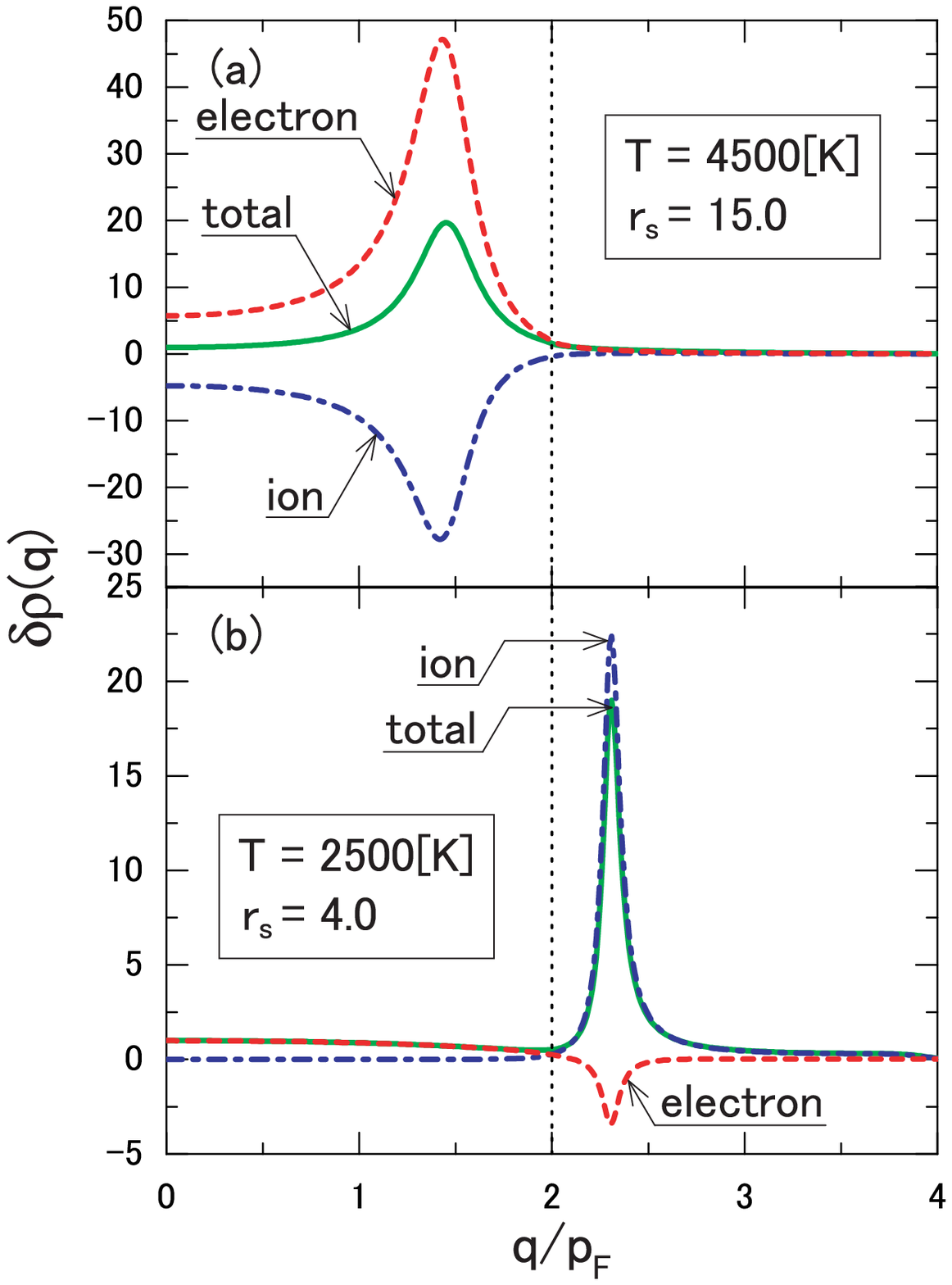}
\end{center}
\caption{Fourier transforms of the electronic, ionic and total charge densities 
induced by a negative point charge at the origin, which are indicated by the 
broken, chain and solid curves, respectively, (a) for $T = 4500$[K] and $r_s 
= 15.0$ and  (b) for $T = 2500$[K] and $r_s = 4.0$.}
\label{fig: induced charges}
\end{figure}

%%%%%%%%%%%%%%%%%%%% Paragraph 5-2: Induced charges    %%%%%%%%%%%%%%%%%%%%%%%%%
To obtain a deeper insight into this charge instability in the dilute fluid 
phase of $r_s > 12.3$, let us consider the charge-density modulation induced 
by a negative point charge put into the origin of the system. This charge-density 
modulation is given by $\delta \rho ({\bi q})$ $=$ $- \chi_{ZZ}({\bi q},0) 
v({\bi q})$ in the linear response theory, which is a total of $\delta 
\rho_{\rm e}({\bi q})$ $=$ $- Z_{\rm e}({\bi q}) \sum_{\alpha} 
\chi_{{\rm e}\alpha}({\bi q},0) Z_{\alpha}({\bi q}) v({\bi q})$ and 
$\delta \rho_{\rm i}({\bi q})$ $=$ $- Z_{\rm i}({\bi q}) \sum_{\alpha} 
\chi_{{\rm i}\alpha}({\bi q},0) Z_{\alpha}({\bi q}) v({\bi q})$ with 
$\delta \rho_{\rm e}({\bi q})$ and $\delta \rho_{\rm i}({\bi q})$ being 
electronic and ionic induced charges, respectively. 
From \eref{sec.3: fluctuation-dissipation theorem}, 
\eref{sec.3: electron-electron response function} 
and \eref{sec.3: electron-ion response function}, we can write down
\begin{eqnarray}
\delta \rho_{\rm e}({\bi q}) &= 
\left( 1 - \frac{1}{\varepsilon({\bi q})} \right)
\left( 1 - \frac{Z_{\rm i}({\bi q})^2 v({\bi q})}{\varepsilon ({\bi q})}
n S_{\rm ii}({\bi q})/T \right)\,,
\\
\delta \rho_{\rm i}({\bi q}) &= \frac{Z_{\rm i}({\bi q})^2 v({\bi q})}
{\varepsilon ({\bi q})} n S_{\rm ii}({\bi q})/T\,.
\end{eqnarray}
\Fref{fig: induced charges} presents $\delta \rho_{\rm e}({\bi q})$ 
(the broken curve), $\delta \rho_{\rm i}({\bi q})$ (the chain curve) and 
$\delta \rho ({\bi q})$ (the solid curve) at two typical points in the 
$r_s$-$T$ plane marked by (a) the circle with dot and (b) the triangle with 
dot in \fref{fig: phase diagram}; we are interested in the charge-density 
modulations for low densities in the former, but for comparison we also show 
those for high densities in the latter which indicate a well-known charge 
instability related to the transition from liquid to solid. In both cases 
of (a) and (b), the charge neutrality condition $\delta \rho ({\bf 0}) = 1$ 
is satisfied by the fact that the external point charge is completely screened 
by the electrons ($|\delta \rho_{\rm e} ({\bf 0})|$ $>$ $|\delta 
\rho_{\rm i} ({\bf 0})|$), although the ions are highly compressible 
($|\delta \rho_{\rm i} ({\bf 0})|$ $\simeq$ $5$) in (a), while almost 
incompressible ($|\delta \rho_{\rm i} ({\bf 0})|$ $\simeq$ $0$) in (b). 

A couple of notable differences exist between (a) and (b) on this charge 
instability: (1) The ``charge-ordering" vector $Q$ as defined by 
the peak position in $\delta \rho ({\bi q})$ is less than $2 p_{\rm F}$, 
the diameter of the Fermi surface of electrons, in (a) but larger than 
$2 p_{\rm F}$ in (b); (2) the charge instability is driven by the electrons, 
while the ions only screen insufficiently the electronic ``charge ordering" 
in (a) because of $|\delta \rho_{\rm e} ({\bi Q})|$ $>$ 
$|\delta \rho_{\rm i} ({\bi Q})|$, but it is done by the ions with 
partially-screening electrons in (b) because of $|\delta 
\rho_{\rm i} ({\bi Q})|$ $>$ $|\delta \rho_{\rm e} ({\bi Q})|$. 

%%%%%%%%%%%%%%%%%%%%%% Paragraph 5-3: electronic transport    %%%%%%%%%%%%%%%%%%
These differences imply that the charge instability for low densities has a 
significant influence on the electronic transport, while that for high 
densities does not. It is clear that the transition from liquid to solid 
keeps the system metallic, while it makes the translation symmetry broken; 
this can be understood by the fact that the charge instability for high 
densities is due to the localization of ions with the charge-ordering vector 
${\bi Q}$, which corresponds to the reciprocal lattice vector of the solid 
with a definite direction, and that the Fermi surface of electrons remains 
almost unchanged since $Q > 2 p_{\rm F}$. On the other hand, the charge 
instability for low densities is characterized by $Q < 2 p_{\rm F}$ and then 
electrons on the Fermi surface connected by ${\bi Q}$ have strong scattering 
amplitudes with a tendency to localization, leading at least to the 
decrease in the electronic conductivity. Particularly in a fluid or gaseous 
phase with the translation symmetry (in a different manner from 
incommensurate-charge-density-wave formation in a solid), 
the entire Fermi surface may be seriously affected by ${\bi Q}$ with randomly 
distributed directions to preserve its rotational invariance, suggesting a 
new route to understanding of the metal-insulator transition.       

\section{Conclusion}
We have investigated the static response functions of the fluid alkali metal 
by applying a perturbation method with respect to the electron-ion interaction 
to the first-principles Hamiltonian and found a charge instability in the 
dilute fluid phase above the liquid-gas critical point in the mean-field 
approximation based on the WCA separation. This charge instability is due 
to electronic density fluctuations at a finite wave vector $Q$ $<$ $2p_F$, 
implying its intimate connection with the metal-insulator transition 
in fluid alkali metals.

\ack
%%%%%%%%%%%%%%%%%%%%%%%%%%     Acknowledgements       %%%%%%%%%%%%%%%%%%%%%%%%%%
This work is partially supported by a Grant-in-Aid for Scientific 
Research in Priority Areas (No.17064004) of MEXT, Japan.

%%%%%%%%%%%%%%%%%%%%%%%%%%%%%%%%%%%%%%%%%%%%%%%%%%%%%%%%%%%%%%%%%%%%%%%%%%%%%%%%

%%%%%%%%% Figure 1:  WCA separation        %%%%%%%%%%%%%.

%%%%%%%%% Figure 2: MFA vs MCS         %%%%%%%%%%%%%.

%%%%%%%%% Figure 3: MFA vs MCS         %%%%%%%%%%%%%.

%%%%%%%%% Figure 4: Phase diagram         %%%%%%%%%%%%%.

%%%%%%%%% Figure 5: Induced charges         %%%%%%%%%%%%%.
%%%%%%%%%%%%%%%%%%%%%%%%%%%%%%%%%%%%%%%%%%%%%%%%%%%%%%%%%%%%%%%%%%%%%%%%%%%%%%%%


\section*{References}
\begin{thebibliography}{10}
\bibitem{Mott90} Mott N F 1990 {\it Metal-Insulator Transitions} 
(London; New York: Taylor and Fransis) Chapter~10
\bibitem{Hensel98} Hensel F 1998 {\it Phil. Trans. R. Soc. Lond.} A {\bf 356} 97 
\bibitem{Mott49} Mott N F 1949 {\it Proc. Phys. Soc.} A {\bf 62} 416
\bibitem{LZ43} Landau L D and Zeldovich G 1943 {\it Acta Physichim. VSSR} 
{\bf 18} 194; 1965 {\it Collected papers of L. D. Landau} ed D Ter Haar 
(Oxford: Pergamon) p~380
\bibitem{HM05} Hansen J P and McDonald I R 2006 {\it Theory of Simple Liquids}
 (London: Elsevier), Fig. 1.1 in p. 2
\bibitem{AS78} Ashcroft N W and Stroud D 1978 {\it Solid State Phys}. {\bf 33} 1 
\bibitem{WH73} Watabe M and Hasegawa M 1973 {\it Properties of Liquid Metals} 
ed S Takeuchi (London: Taylor and Francis) p~133
\bibitem{MCS95} Moroni S, Ceperley D M and Senatore G 1995 \PRL {\bf 75} 689
\bibitem{WCA71} Weeks J D, Chandler D and Andersen H C 1972 \JCP {\bf 54} 5237
\bibitem{PY58} Percus J K and Yevick G J 1958 \PR {\bf 110} 1
\bibitem{Thiele63} Thiele E 1963 \JCP {\bf 39} 474 
\bibitem{Wertheim63} Wertheim M S 1963 \PRL {\bf 10} 321
\bibitem{MTI07}Matsuda K, Tamura K and Inui M 2007 \PRL {\bf 98} 096401
%%%%%%%%%%%%%%%%%%%%%%%%%%%%%%%%%%%%%%%%%%%%%%%%%%%%%%%%%%%%%%%%%%%%%%%%%%%%%%%%
\end{thebibliography}
\end{document}